\documentclass[a4paper, 12pt]{article}       
\usepackage[latin1]{inputenc}
\usepackage{alltt,graphicx,amssymb,epsfig}
\usepackage{amsmath,mathrsfs,subfigure}

\usepackage{color}

\def \tr{\mbox{tr\hskip 1pt}}

\def\W{\mathcal{W}}
\def\I{I^*}

\renewcommand{\vec}[1]{\boldsymbol{#1}}

\def \tr{\mbox{tr\hskip 1pt}}

\def\emm{\vec{M}\cdot\mathbf{e}\vec{M}}
\def\empmp{\vec{M}'\cdot\mathbf{e}\vec{M}'}
\def\emmp{\vec{M}\cdot\mathbf{e}\vec{M}'}

\begin{document}

\title{On anisotropic elasticity and questions concerning its Finite Element implementation
}

\author{Luigi Vergori$^1$, Michel Destrade$^{1,2}$,\\ Patrick McGarry$^3$, Ray W. Ogden$^4$\\[12pt]
$^1$School of Mathematics, Statistics \& Applied Mathematics\\
National University of Ireland Galway, Ireland, \\[12pt]
$^2$School of Mechanical \& Materials Engineering,\\
University College Dublin, Belfield, Dublin 4, Ireland,\\[12pt]
$^3$Mechanical \& Biomedical Engineering,\\
National University of Ireland Galway, Ireland,\\[12pt]
$^4$School of Mathematics \& Statistics,\\
 University of Glasgow, Scotland}
\date{}

\maketitle

\begin{abstract}

We give conditions on the strain-energy function of nonlinear ani\-so\-tro\-pic hyperelastic materials that ensure compatibility with the classical linear theories of anisotropic elasticity. We uncover the limitations associated with the volumetric deviatoric separation of the strain energy used, for example, in many Finite Element (FE) codes in that it does not fully represent the behavior of anisotropic materials in the linear regime. This limitation has important consequences. We
show that, in the small deformation regime, a FE code based on the volumetric-deviatoric separation assumption predicts that a sphere made of a compressible anisotropic material deforms into another sphere under hydrostatic pressure loading, instead of the expected ellipsoid. For finite deformations, the commonly
adopted assumption that fibres cannot support compression is incorrectly implemented in current FE codes and leads to the unphysical result that under
hydrostatic tension a sphere of compressible anisotropic material deforms into a larger sphere.

\end{abstract}

\noindent
\textbf{keywords:} Anisotropic elasticity; Nonlinear hyperelasticity; Finite Elements; Deviatoric--volumetric decoupling

\section{Introduction}\label{Introduction}

\begin{figure}[!ht]
\centerline{
a)\includegraphics[width=0.34\textwidth]{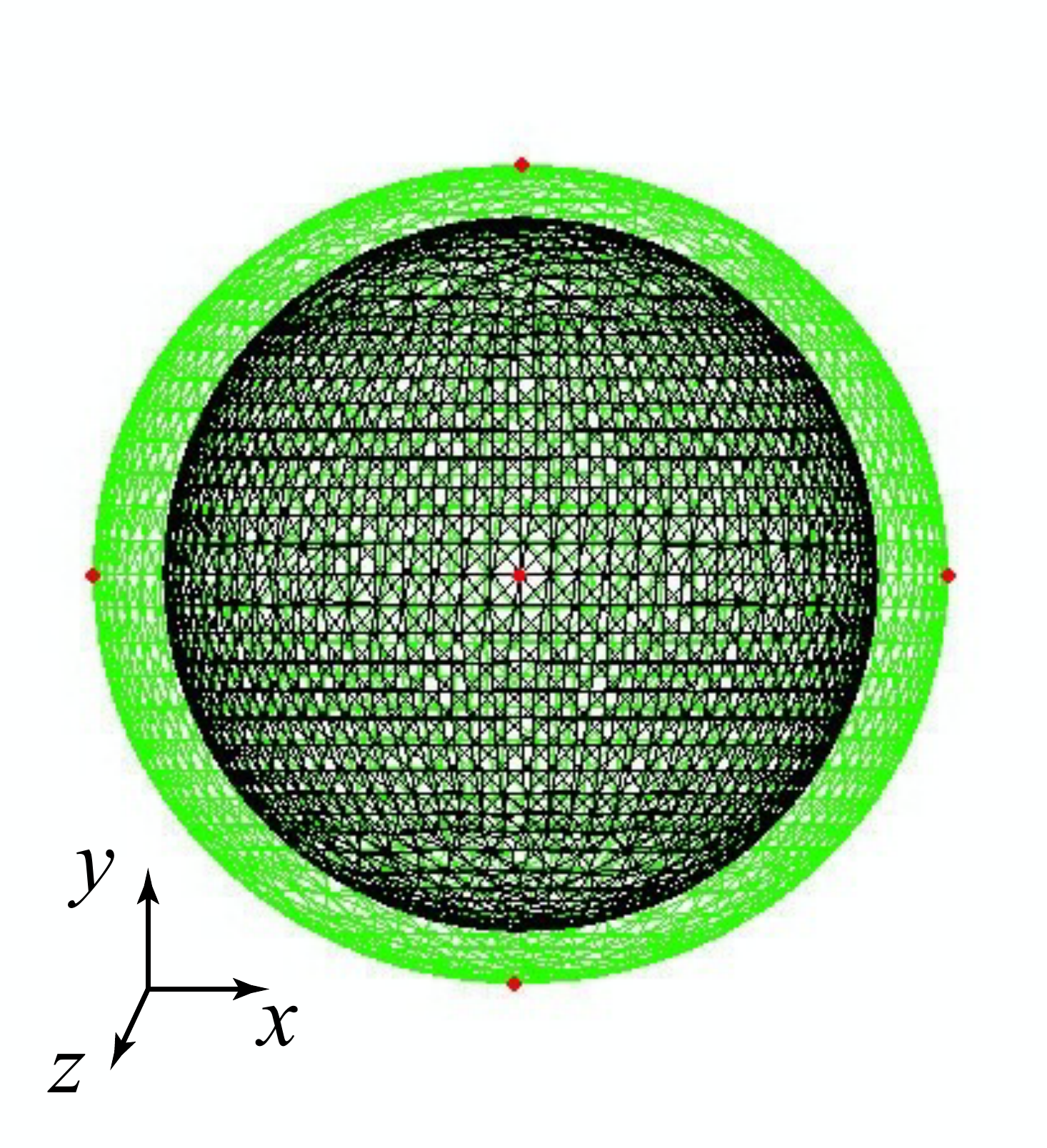}
b)\includegraphics[width=0.44\textwidth]{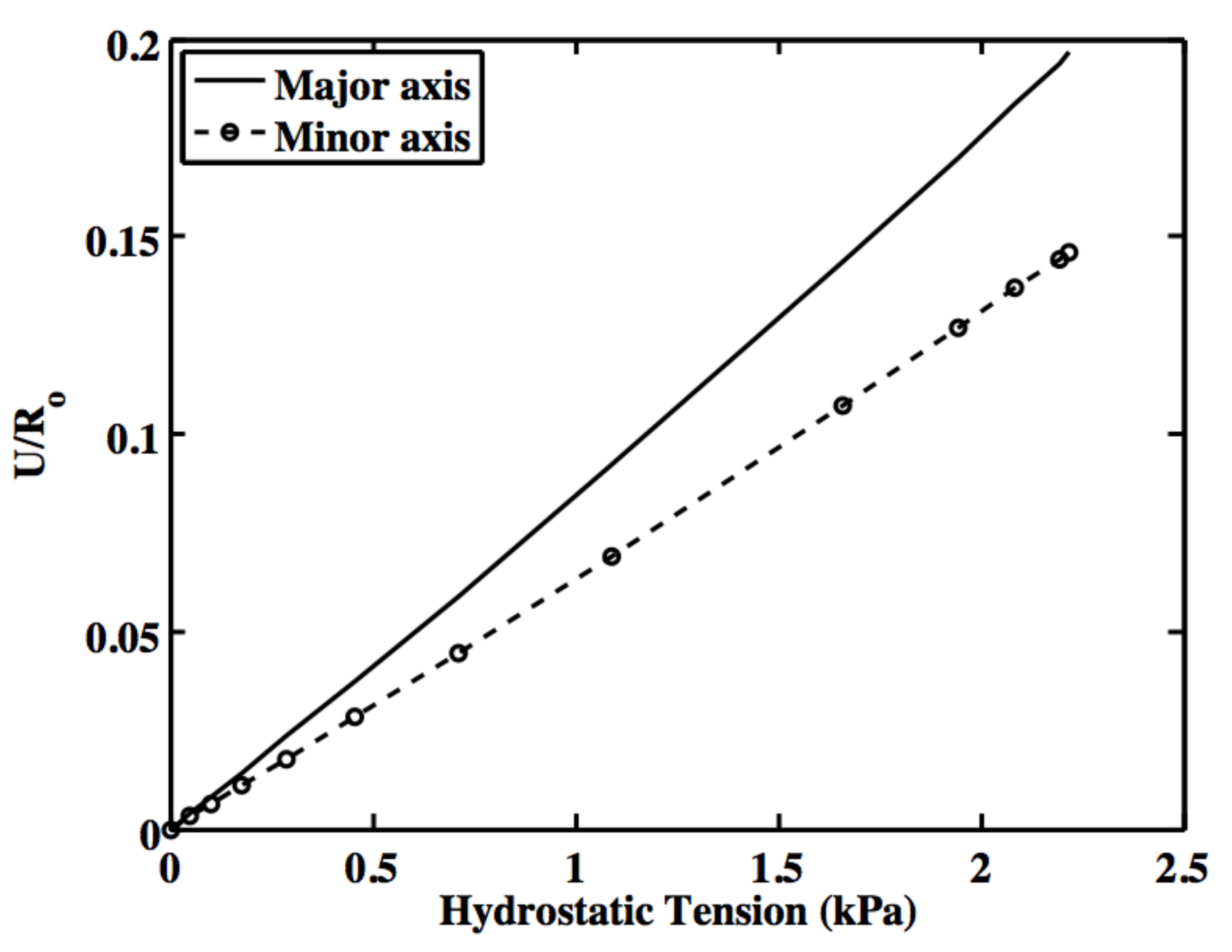}
}
\caption{
{\small In Abaqus\textsuperscript{\textregistered}, \emph{(a)} a sphere with the linearly elastic constants of paratellurite (transversely isotropic Tellurium dioxide TeO$_2$; see, for example, \cite{RoD84}) deforms into an ellipsoid (green) under hydrostatic tension; \emph{(b)} shows the magnitude of  displacement $U$ (normalized by the radius of the undeformed sphere $R_o$) of six points (highlighted in red in \emph{(a)}) on the extremities of the major axis \color{black} (along the $x$-axis) \color{black} and minor axes \color{black} (along the $y$- and the $z$-axes) \color{black} of the ellipsoid as a function of the applied uniform surface tension (in kPa).}}
\label{fig-TeO2-tension}
\end{figure}

Intuitively, one would expect a sphere made of a homogeneous anisotropic elastic material to deform into an ellipsoid when subjected to hydrostatic stress.
This is quite simple to prove for the case of a linearly elastic, transversely isotropic material, with isotropy in the $(x_1,x_2)$ plane, say.
Using the Voigt notation, the five independent elastic constants are $c_{11}, c_{12}, c_{13}, c_{33}, c_{44}$.
When we apply a hydrostatic stress $\boldsymbol{\sigma} = -p\mathbf{I}$, where $p$ is the pressure (when positive) or tension (when negative), the non-zero components of the strain $\mathbf e$ are easily found to be
\begin{equation} \label{hydrostat}
\left.\begin{array}{ll}
 e_{11} = e_{22} = -p \dfrac{c_{33}-c_{13}}{(c_{11}+c_{12})c_{33} - 2 c_{13}^2}, \\
 [4mm]
 e_{33} = -p \dfrac{c_{11}  + c_{12} -2c_{13}}{(c_{11}+c_{12})c_{33} - 2 c_{13}^2},
 \end{array}\right.
\end{equation}
(see for instance Musgrave \cite{Musg70}), and these are clearly unequal in general.  It should be noted in passing that the denominator here is never zero since (provided the material is compressible \cite{DeMT02}) the stiffness matrix is required to be positive definite. This is confirmed numerically in the simulation of Figure \ref{fig-TeO2-tension}, where we used Abaqus\textsuperscript{\textregistered} to deform a homogeneous sphere of \emph{linearly} elastic anisotropic material into an ellipsoid by applying a hydrostatic tension  (i.e. a uniform tensile loading $(p<0)$ is applied to the surface of the sphere). Here the applied tension is increased so that the final computed deformation shown in Figure \ref{fig-TeO2-tension}(a) exceeds the small strain regime to produce a ``visible'' deformation for illustrative purposes. However, it is worth noting that the ellipsoidal shape emerges from the start of the small deformation regime, as exemplified by the immediate difference between the deformed lengths of the major axis and the minor axes upon the application of the hydrostatic tensile loading, as shown in Figure \ref{fig-TeO2-tension}(b). For an incompressible material, by contrast, an anisotropic sphere does not deform at all under a hydrostatic stress (this is shown theoretically in the nonlinear context in Section \ref{Hydrostatic stress}).

\begin{figure}[!ht]
\centerline{
a)\includegraphics[width=0.4\textwidth]{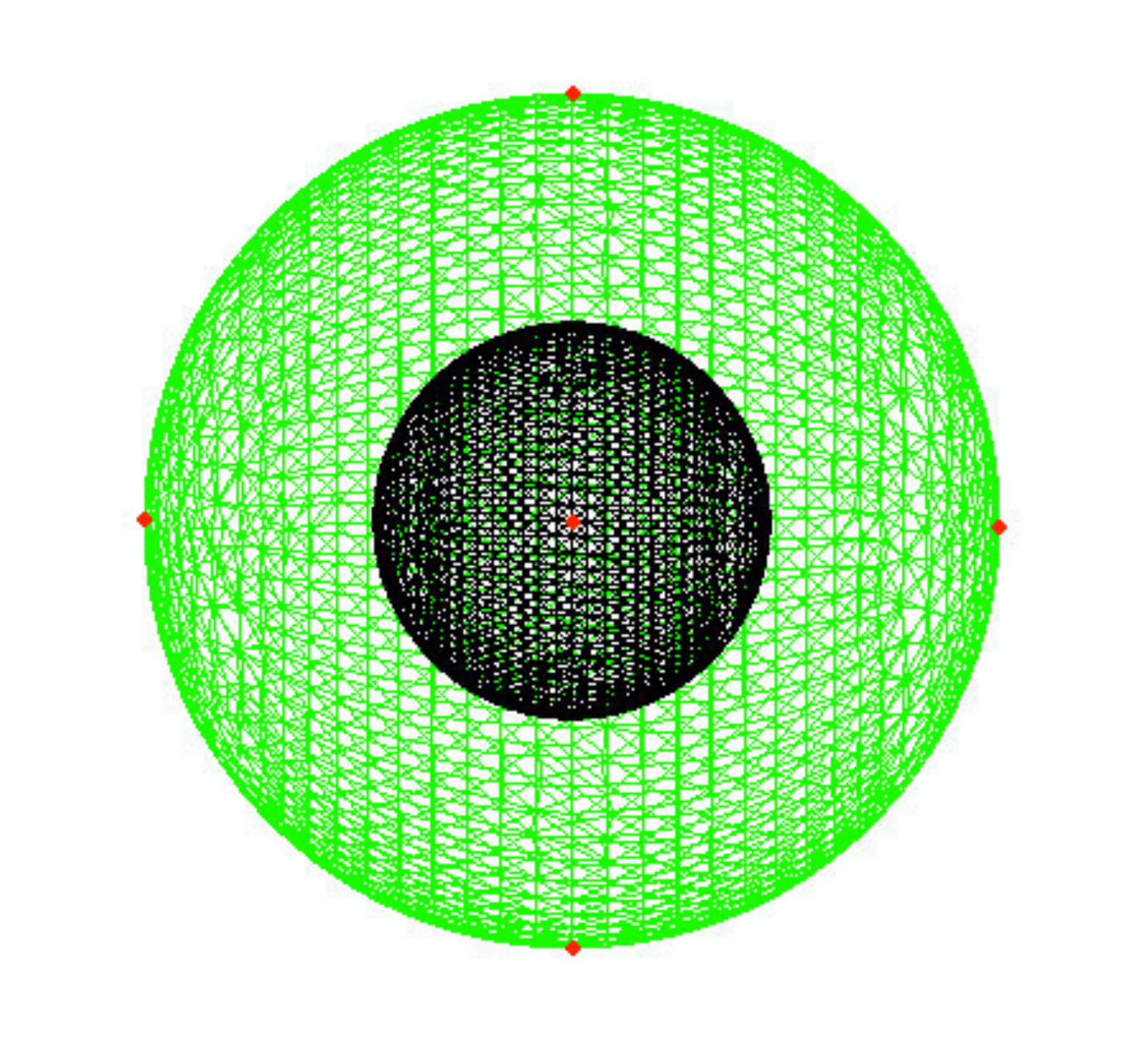}
b)\includegraphics[width=0.34\textwidth]{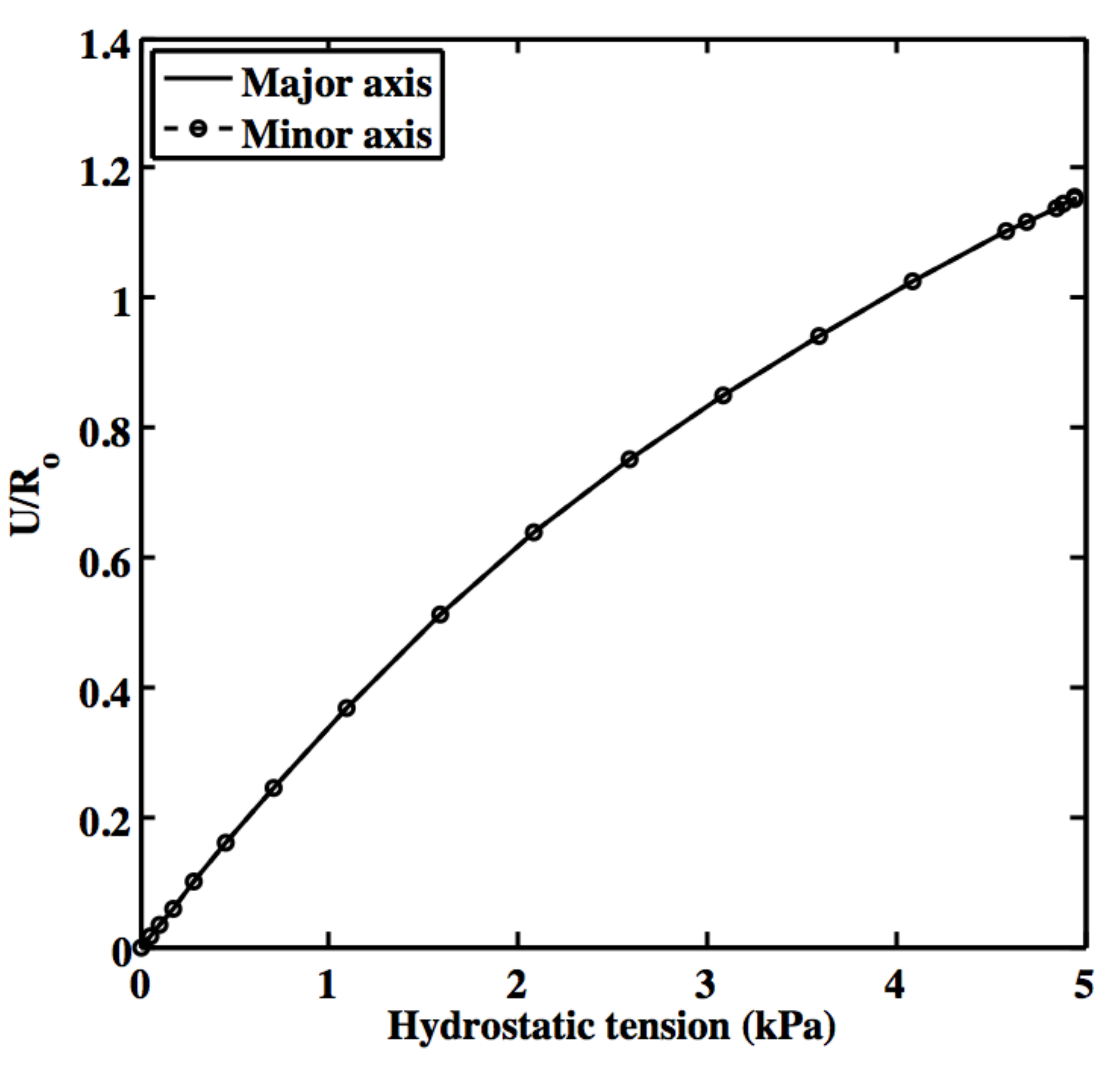}
}
\caption{
{\small In Abaqus\textsuperscript{\textregistered}, \emph{(a)} a sphere (black) modelled by the ``Holzapfel--Gasser--Ogden'' model deforms into a larger sphere (green) under hydrostatic tension.
In the simulation, the loading produces a large deformation and the spherical shape remains, as attested by  \emph{(b)} which displays the non-dimensionalized displacements of the major and minor axes end-points as functions of the hydrostatic stress (in kPa).}}
\label{fig-holzapfel-tension}
\end{figure}

However, when we use one of the \emph{nonlinear} {hyperelastic} models of ani\-so\-tro\-pic materials instead in Abaqus\textsuperscript{\textregistered} to perform the same numerical experiment, we find that a sphere is transformed into a sphere of larger diameter, not an ellipsoid, as illustrated in Figure \ref{fig-holzapfel-tension}. In this paper we show that this unphysical prediction is explained by the {\color{black}way that certain models of nonlinear anisotropic elasticity are implemented in Finite Element (FE) codes, as a result of which they are} unable to predict, either in the infinitesimal (linear) limit or in the finite deformation regime, the correct ellipsoidal shape due to hydrostatic loading.  In this paper we investigate the theory underlying this problem.

The paper is organized as follows.  In Section \ref{Inconsistency with transverse isotropy} we recall the equations of compatibility which ensure that a nonlinear {hyper}elastic material reinforced with one family of parallel fibres behaves like a transversely iso\-tropic solid in the infinitesimal limit \cite{o-m}. These are then re-cast in terms of the volumetric/deviatoric formulation of nonlinear {hyper}elasticity, and we specialize them to the case of a strain-energy function that is decoupled into the sum of a volumetric part and a deviatoric part.  In this case we find  that the number of independent elastic constants is reduced from five to four and thus the full transverse isotropy of linear elasticity is not captured.  In particular, this specialization implies that a sphere deforms into a sphere under hydrostatic stress, i.e. that $e_{11} = e_{22} = e_{33}$ in \eqref{hydrostat}.

In Section \ref{Inconsistency with orthotropy} we extend the transversely isotropic result to the case of a material reinforced to two families of orthogonal fibres, for which the constitutive equations reduce to a specialization of orthotropic linear elasticity with only seven elastic constants instead of the nine required for full orthotropy.  We also give results for the general case of a material reinforced with two families of (not necessarily orthogonal) fibres, for which the 13 elastic constants associated with this monoclinic case reduce to 10, but the details are relegated to Appendix A. This covers the case where the two families are made of fibres with the same mechanical properties, as implemented in the so-called ``Holzapfel--Gasser--Ogden'' model \cite{holz1}.

In Section \ref{Numerical experiments} we give the details of the simulation conducted in Abaqus\textsuperscript{\textregistered} using its version of the latter model.  As predicted from the analysis of the preceding sections, Abaqus\textsuperscript{\textregistered} deforms a sphere into another sphere in the small deformation regime. Surprisingly, it does so even in the large deformation regime, even though, as we establish in Section \ref {Hydrostatic stress}, the theory shows that the linear result does not carry over to the nonlinear regime: for a compressible anisotropic material a hydrostatic stress does not yield a uniform dilatation in nonlinear {hyper}elasticity, whether or not the strain energy is decoupled into volumetric/deviatoric parts.
Here the unphysical behaviour of a sphere turning into a sphere is due to the way that {\color{black}some anisotropic models are} implemented in Abaqus\textsuperscript{\textregistered}. {\color{black} In particular, models for which the anisotropic contribution is active} only when the fibres are in extension, i.e. when $\lambda>1$, where $\lambda$ is the stretch in the fibre direction.
Current FE codes {\color{black} effectively express} this condition incorrectly as $\lambda^*>1$ instead, where $\lambda^*$ is the \emph{deviatoric} stretch in the fibre direction.
A dramatic consequence of this difference is that the anisotropic contribution is never called upon under a hydrostatic stress loading.  Since more complex deformations computed for soft tissue by FE codes typically include a hydrostatic part, it follows that the validity of such FE calculations must be viewed with caution.

In Section \ref{why}, we show why, on the basis of the assumptions made in FE implementations, a sphere becomes a sphere under hydrostatic  tensile loading contrary to physical expectation.  Finally, as already mentioned, results for monoclinic material symmetry are presented in Appendix A.


\section{Transverse isotropy}
\label{Inconsistency with transverse isotropy}


For solids with one family of parallel fibres, Spencer  \cite{Spen72} showed that in general, the strain-energy function $W$ is a function of three isotropic strain invariants ($I_1, I_2, I_3$) and two anisotropic invariants ($I_4,I_5)$, which together form a complete set of independent invariants for transversely isotropic elastic materials. Hence
\begin{equation}
W = W(I_1,I_2,I_3,I_4,I_5),
\end{equation}
where
 \begin{align}\label{I1I5}
& I_1 = \tr\mathbf{C}, \quad  I_2 = \frac{1}{2}\left[\left(\tr\mathbf{C}\right)^2  -\tr\left(\mathbf{C}^2\right)\right], 
\quad I_3 = \det \mathbf{C}, \notag\\
&  I_4 =  \vec{M} \cdot\mathbf{C} \vec{M}, \quad  I_5 = \vec{M}\cdot \mathbf{C}^2 \vec{M},
\end{align}
 $\mathbf{F}$ is the deformation gradient, $\mathbf{C} = \mathbf{F}^\mathrm{T} \mathbf{F}$ is the right Cauchy--Green deformation tensor and $\vec{M}$ is a unit vector aligned with the fibres in the reference configuration, which is assumed to be stress free.

Here we give connections between the first and second derivatives of $W$ with respect to the invariants, evaluated in the undeformed configuration, and the five constants of transversely isotropic linear elasticity as a prelude to considering the implications of the volumetric/deviatoric separation.

First, the Cauchy stress $\boldsymbol{\sigma}$ is given by
\begin{multline}
J\boldsymbol{\sigma}=2W_1\mathbf{B}+2W_2(I_1\mathbf{B}-\mathbf{B}^2)+2I_3W_3\mathbf{I}\\
+2W_4\vec{m}\otimes\vec{m}+2W_5(\vec{m}\otimes\mathbf{B}\vec{m}+\mathbf{B}\vec{m}\otimes\vec{m}),\label{Cauchy1}
\end{multline}
where $J = \det \mathbf{F}=I_3^{1/2}$ is the dilatation, $\mathbf{B} = \mathbf{FF}^\mathrm{T}$ is the left Cauchy--Green deformation tensor, {\color{black}$\mathbf{I}$ is the identity tensor}, $\vec{m} = \mathbf{F}\vec{M}$,  and $W_k = \partial W/\partial I_k,\,k=1,2,\ldots,5$.

A useful alternative way to express the constitutive law is in terms of the invariants of the \emph{distortional part} of the deformation \cite{holz1,ogden78}.  This is defined by
\begin{equation}
\mathbf{F}^*=J^{-1/3}\mathbf{F}, \quad  \text{so that} \quad \det\mathbf{F}^*=1,
\end{equation}
indicating that no volume changes occurs for the part of the deformation described by $\mathbf{F}^*$.
The corresponding distortional right and left Cauchy--Green tensors are
\begin{equation}
\mathbf{C}^*=J^{-2/3}\mathbf{C},\qquad \mathbf{B}^*=J^{-2/3}\mathbf{B},
\end{equation}
by means of which the associated deviatoric invariants
{\color{black}\begin{eqnarray}
 \I_1 &=&J^{-2/3} I_1, \quad  \I_2 = J^{-4/3}I_2,\notag\\[1ex]
    \I_4 &=& J^{-2/3} I_4, \quad  \I_5 = J^{-4/3}I_5,
\label{I1I5star}
\end{eqnarray}}
are constructed.

We now take the energy function to depend on the independent invariants $\I_1,\I_2, I_3,\I_4,\I_5$ and denote it by $W^*$.
Then, equation \eqref{Cauchy1} can be rewritten as
\begin{align}
J\boldsymbol{\sigma} =& 2W^*_1(\mathbf{B}^*-\tfrac{1}{3}I_1^*\mathbf{I})+2W^*_2(I^*_1\mathbf{B}^*-\mathbf{B}^{*2}-\tfrac{2}{3}I_2^*\mathbf{I})\notag\\
&+2I_3W^*_3\mathbf{I}
+2W^*_4(\vec{m}^*\otimes\vec{m}^*-\tfrac{1}{3}I_4^*\mathbf{I})\notag\\
&+2W^*_5(\vec{m}^*\otimes\mathbf{B}^*\vec{m}^*+\mathbf{B}^*\vec{m}^*\otimes\vec{m}^*-\tfrac{2}{3}I_5^*\mathbf{I}),\notag
\end{align}
where $\vec{m}^*= \mathbf{F}^* \vec{M} = J^{-1/3}\vec{m}$.
An immediate consequence is that
\begin{equation}
\tfrac{1}{3}\tr\boldsymbol{\sigma}=2JW^*_3\equiv W^*_J,
\end{equation}
which is a special case of a general result for hyperelastic materials given in \cite{ogden78}.

In the (undeformed) reference configuration, $\mathbf{F} = \mathbf{C} = \mathbf{I}$ and the invariants take the values
\begin{equation}
I_1= I_2=3, \quad I_3 =1, \quad   I_4 = I_5=1,\label{invariants-in-ref-config}
\end{equation}
and thus
\begin{equation}
J=1, \quad \I_1=\I_2=3, \quad \I_4=\I_5=1.
\end{equation}
For convenience, the energy function is assumed to be zero in the undeformed state, and {\color{black} we also assume that the stress vanishes} there.
This leads to  (see Merodio and Ogden \cite{o-m})
\begin{equation}
W=0,\quad
W_1+ 2W_2+ W_3= 0, \quad W_4+2W_5= 0,
\label{zero}
\end{equation}
wherein all the terms are evaluated for \eqref{invariants-in-ref-config}.
Consistency with the classical linear theory of transversely isotropic elasticity is achieved when the additional conditions
\begin{align}\label{general}
& W_1 + W_2 +W_5 =c_{44}/2, \nonumber \\[0.5ex]
&W_2 +W_3 =\frac{c_{12} -c_{11}}{4}, \nonumber\\[0.5ex]
&W_{11} + 4W_{12} +4W_{22} +2W_{13} +4W_{23} +W_{33} = \frac{c_{11}}{4}, 	\nonumber \\[0.5ex]
&W_{14} +2W_{24} +2W_{15} +W_{34} +4W_{25} + 2W_{35} =\frac{c_{13}- c_{12}}{4},	\notag \\[0.5ex]
&W_{44} +4W_{45} +4W_{55} +2W_5 =\frac{c_{33} -c_{11}+ 2c_{12} -2c_{13}}{4} 
\end{align}
are satisfied  \cite{o-m}, where again the derivatives of $W$ are evaluated in the reference configuration.
Here the \emph{five} independent elastic constants $c_{11}$,..., $c_{44}$ are given in the standard Voigt notation with the $x_3$ coordinate direction corresponding to the axis of symmetry aligned with the unit vector $\vec{M}$.

The corresponding, equivalent, results in terms of $W^*$ are simply
\begin{align}\label{W*}
& W^*= 0,\quad W^*_1+W^*_2=\tfrac{1}{4}(c_{11}-c_{12}),\quad W_3^*=0,\nonumber \\[0.5ex]
& W^*_4=-2W^*_5=\tfrac{1}{2}(c_{11}-c_{12})-c_{44},
\end{align}
and
\begin{align} \label{W*TI}
& W^*_{33} = \tfrac{1}{36}(2c_{11}+2c_{12}+c_{33}+4c_{13}),\nonumber \\[0.5ex]
& W^*_{34}+2W^*_{35} = \tfrac{1}{12}(c_{13}-c_{12}+c_{33}-c_{11}),\nonumber \\[0.5ex]
& W^*_{44}+4W^*_{45}+4W^*_{55} = \tfrac{1}{4}(c_{11}+c_{33}-2c_{13})-c_{44},
\end{align}
all evaluated in the reference configuration.

The hyperelastic models implemented in many Finite Element codes rely on an additive separation of the strain-energy function $W$ into a \emph{volumetric part} and a \emph{deviatoric part}.
For solids with one family of parallel fibres, $W$ is thus \emph{assumed} to be of the form
\begin{align}\label{assumption1}
\nonumber
W(I_1,I_2,I_3,I_4,I_5)&=W^*(\I_1,\I_2,I_3,\I_4,\I_5)\\
&= f(I_3)+\W(\I_1,\I_2,\I_4,\I_5),
\end{align}
where $f$, the volumetric part, is a function of $I_3$ only and $\W$, the deviatoric part, is a function of the deviatoric isotropic ($\I_1$, $\I_2$) and anisotropic ($\I_4$, $\I_5$) invariants only.
Separation into  volumetric and deviatoric parts is done for instance for the so-called ``Fung'' and ``Holzapfel--Gasser--Ogden'' models in Abaqus\textsuperscript{\textregistered} \cite{abaqus}, the ``Holzapfel'' model in ANSYS\textsuperscript{\textregistered} \cite{ansys}, the ``Transversely Isotropic Hyperelastic'', ``Fung Orthotropic", ``Tension--Compression Nonlinear Orthotropic'' models in FEBio \cite{febio}, and the ``Orthotropic Effects" model in ADINA\textsuperscript{\textregistered} \cite{adina}.

We  now substitute the special form of strain-energy function \eqref{assumption1} into the general equations \eqref{W*}, \eqref{W*TI}.  The only one to give anything essentially new is equation \eqref{W*TI}$_2$.
Its left-hand side is clearly zero when the decoupling is enforced, which gives
\begin{equation} \label{special0}
c_{12}-c_{13}= c_{33}-c_{11}.
\end{equation}
If this relation were to hold, then there would be only 4 independent elastic constants and the material would not be fully transversely isotropic in the linear regime.  It follows that materials with one family of parallel fibres for which the strain-energy function can be decomposed additively as \eqref{assumption1} do not behave like general transversely isotropic solids when subject to infinitesimal deformations. {\color{black}This is not surprising since \eqref{assumption1} represents a specialization of the transversely isotropic theory, and it has not been claimed otherwise in the literature.}
\color{black}However\color{black},  as shown by Musgrave \cite{Musg70}, if the elastic constants of a transversely isotropic solid were to obey equation \eqref{special0}, then the material would contract or dilate \emph{uniformly} under either a compressive or a tensile hydrostatic stress.
This is clearly seen by comparing equations \eqref{hydrostat} and \eqref{special0}.

This conflict puts simulations based on that decomposition into perspective, because the resulting material will not behave as expected physically.
These results carry over to the case of a solid with two families of fibres, as we show in the next section and Appendix A.


\section{Orthotropy}
\label{Inconsistency with orthotropy}


A material reinforced with two families of parallel fibres possesses only one plane of symmetry (the plane containing all the fibres) in general, and is thus of monoclinic symmetry.  In the linear theory a monoclinic solid has 13 independent elastic constants.
As shown by Spencer \cite{Spen72}, there are two special cases of materials with two planes of symmetry: (a) when the fibres are at right angles, and (b) when the fibres are all mechanically equivalent.
In those two cases, the constitutive equations should specialize to those of linearly elastic \emph{orthotropic} solids in the infinitesimal limit, which possess \emph{nine} independent elastic constants. Here we extend the results of the previous section to Case (a). Case (b) is lengthier and can be read off from the general monoclinic treatment presented in Appendix A.

When the fibre families are orthogonal, the strain-energy function $W$ depends on 7 invariants \cite{MeOg06}: $I_1$ to $I_5$ defined in
\eqref{I1I5}, and
\begin{equation}
 I_6=  \vec{M}' \cdot \mathbf{C}\vec{M}', \qquad  I_7 = \vec {M}' \cdot \mathbf{C}^2 \vec{M}',
\end{equation}
$\vec{M}'$ being a unit vector in the reference configuration orthogonal to $\vec M$ and  aligned with the second family of fibres.
Without loss of generality, $\vec M$ and $\vec{M}'$ may be taken to be aligned with two orthogonal Cartesian unit vectors, $\vec{e}_2$ and $\vec{e}_3$ say.
In the stress-free reference configuration,  where
\begin{equation}\label{ref-ortho}
I_1=I_2=3,\quad I_3=I_4=I_5=I_6=I_7=1,
\end{equation}
we must have
\begin{equation}\label{zero-ortho}
W_1+2W_2+W_3=0,\quad W_4+2W_5=0, \quad W_6+2W_7=0.
\end{equation}
Furthermore, we find that consistency with the classical linear theory of orthotropic elasticity is achieved when the conditions
\begin{align}\label{consistency conditions perp}
&
W_1+W_2=\tfrac{1}{2}(c_{55}+c_{66}-c_{44}),
\notag \\[8pt]
&
W_5= \tfrac{1}{2}(c_{44}-c_{66}),
\notag \\[8pt]
&
W_7= \tfrac{1}{2}(c_{44}-c_{55}),
\notag \\[8pt]
&
W_{11}+4W_{12}+4W_{22}+2W_{13}+4W_{23}+W_{33}= \tfrac{1}{4} c_{11},
\notag \\[8pt]
&
W_{14}+2W_{24} +2W_{15} + W_{34} + 4W_{25}\notag \\
&\quad+2W_{35}=\tfrac{1}{4} [c_{13}-c_{11}+2(c_{55}+c_{66}-c_{44})],
\notag \\[8pt]
&
W_{16}+2W_{26}+W_{36}+2W_{17}+4W_{27}\notag\\
&\quad+2W_{37}=\tfrac{1}{4}\left[c_{12}-c_{11}+2(c_{55}+c_{66}-c_{44})\right],
\notag \\[8pt]
&
W_{44}+4W_{45}+4W_{55}= \tfrac{1}{4}(c_{11}+c_{33}-2c_{13}-4c_{55}),
\notag \\[8pt]
&
W_{66}+4W_{67}+4W_{77}=\tfrac{1}{4}(c_{11}+c_{22}-2c_{12}-4c_{66}),
\notag \\[8pt]
&
W_{46}+2W_{47}+2W_{56}+4W_{57}\notag\\
&\quad= \tfrac{1}{4}\left[c_{11} - c_{12} +c_{23} - c_{13} - 2(c_{55} + c_{66} - c_{44})\right],
\end{align}
are met, where the derivatives of $W$ are evaluated in the reference configuration.
Here $c_{11}$, $c_{12}$, $c_{13}$, $c_{22}$, $c_{23}$, $c_{33}$, $c_{44}$, $c_{55}$, $c_{66}$ are the nine independent elastic constants in the Voigt notation.

Similarly to \eqref{I1I5star}$_{4,5}$, we introduce the deviatoric invariants
{\color{black}\begin{equation}
  \I_6 = J^{-2/3} I_6, \quad  \I_7 = J^{-4/3} I_7,
\label{I6I7star}
\end{equation}}
and consider the strain-energy function $W^*$ to depend in all generality on the  seven invariants: $$W^* = W^*(\I_1,\I_2,I_3,\I_4,\I_5,\I_6,\I_7).$$
Then the above compatibility equations take the equivalent, but more compact, forms
\begin{align}
& W^*_3=0,\nonumber \\[0.5ex]
& W^*_1+W^*_2=\tfrac{1}{2}(c_{55}+c_{66}-c_{44}),
\nonumber \\[0.5ex]
& W^*_4=-2W^*_5=c_{66}-c_{44},\nonumber \\[0.5ex]
& W^*_6=-2W^*_7=c_{55}-c_{44},
\end{align}
and
\begin{align}
&
W^*_{33}=\tfrac{1}{36}(c_{11}+c_{22}+c_{33}+2c_{12}+2c_{23}+2c_{13}),
\nonumber \\[8pt]
&W^*_{34}+2W^*_{35}=\tfrac{1}{12}(c_{33}+c_{23}-c_{11}-c_{12}),
\nonumber \\[8pt]
&W^*_{36}+2W^*_{37}=\tfrac{1}{12}(c_{22}+c_{23}-c_{11}-c_{13}),
\nonumber \\[8pt]
&W^*_{44}+4W^*_{45}+4W^*_{55}=\tfrac{1}{4}(c_{11}+c_{33}-2c_{13}-4c_{55})
\nonumber \\[8pt]
&W^*_{66}+4W^*_{67}+4W^*_{77}=\tfrac{1}{4}(c_{11}+c_{22}-2c_{12}-4c_{66})
\nonumber \\[8pt]
&W^*_{46}+2W^*_{47}+2W^*_{56}+4W^*_{57}\notag\\
&\quad=\tfrac{1}{4}[c_{11}-c_{12}+c_{23}-c_{13}-2(c_{55}+c_{66}-c_{44})],\label{W*ortho}
\end{align}
all evaluated in the reference configuration where the invariants have the values given in \eqref{ref-ortho} .

Again, we now look at the implications of the volumetric/deviatoric decoupling, in this case in the form
\begin{equation}
W = W^* = f(I_3)+\W(I_1^*,I_2^*,I_4^*,I_5^*,I_6^*,I_7^*).\label{assumption2}
\end{equation}
Clearly, because of the decoupling, the left-hand sides of the second and third equations in \eqref{W*ortho} are zero, which yields the following two conditions on the elastic stiffnesses:
\begin{equation} \label{special1}
c_{23}-c_{12}= c_{11}-c_{33}, \quad
c_{23}-c_{13}=c_{11}-c_{22}.
\end{equation}

If these relations hold, then there are only 7 independent elastic constants and the material is not a general orthotropic material in the linear regime.
In fact, it is then a very special orthotropic material, for which a sphere deforms into a sphere when subject to {compressive or tensile} hydrostatic pressure  {loadings}.
This clearly unphysical result is easily shown by extending the analysis of Musgrave \cite{Musg70} to the present case.

It follows that materials with two orthogonal families of parallel fibres for which the strain-energy function is decomposed additively as in \eqref{assumption2} do not behave like general orthotropic solids when subject to infinitesimal deformations.
Similar results can also be deduced for the other case of orthotropy, when the families of fibres are not necessarily at right angles, but are mechanically equivalent (see Appendix A for general case). This includes the popular Holzapfel--Gasser--Ogden model \cite{holz1} for arteries, for which the original incompressible formulation has been decoupled into a deviatoric/volumetric split in the Abaqus\textsuperscript{\textregistered} and ADINA\textsuperscript{\textregistered} implementations.

To summarize: separating the strain-energy function of a nonlinearly hyperelastic anisotropic material into the sum of a deviatoric part and a volumetric part leads to a model which fails to fully capture  linear anisotropic elasticity in the small deformation regime.
In particular, the decoupling predicts that in the linearized regime, a sphere deforms into another sphere under hydrostatic  loading -- a clearly unphysical behaviour. We now illustrate this prediction with Abaqus\textsuperscript{\textregistered}.


\section{Numerical experiments}
\label{Numerical experiments}


We subjected Abaqus\textsuperscript{\textregistered} to a test in linear anisotropic elasticity (where there is no decoupling of the strain energy between volumetric and deviatoric parts) and a test in nonlinear anisotropic {hyperelasticity} (where the decoupling is implemented by default).

\begin{figure}[!ht]
\centerline{
a)\includegraphics[width=0.22\textwidth]{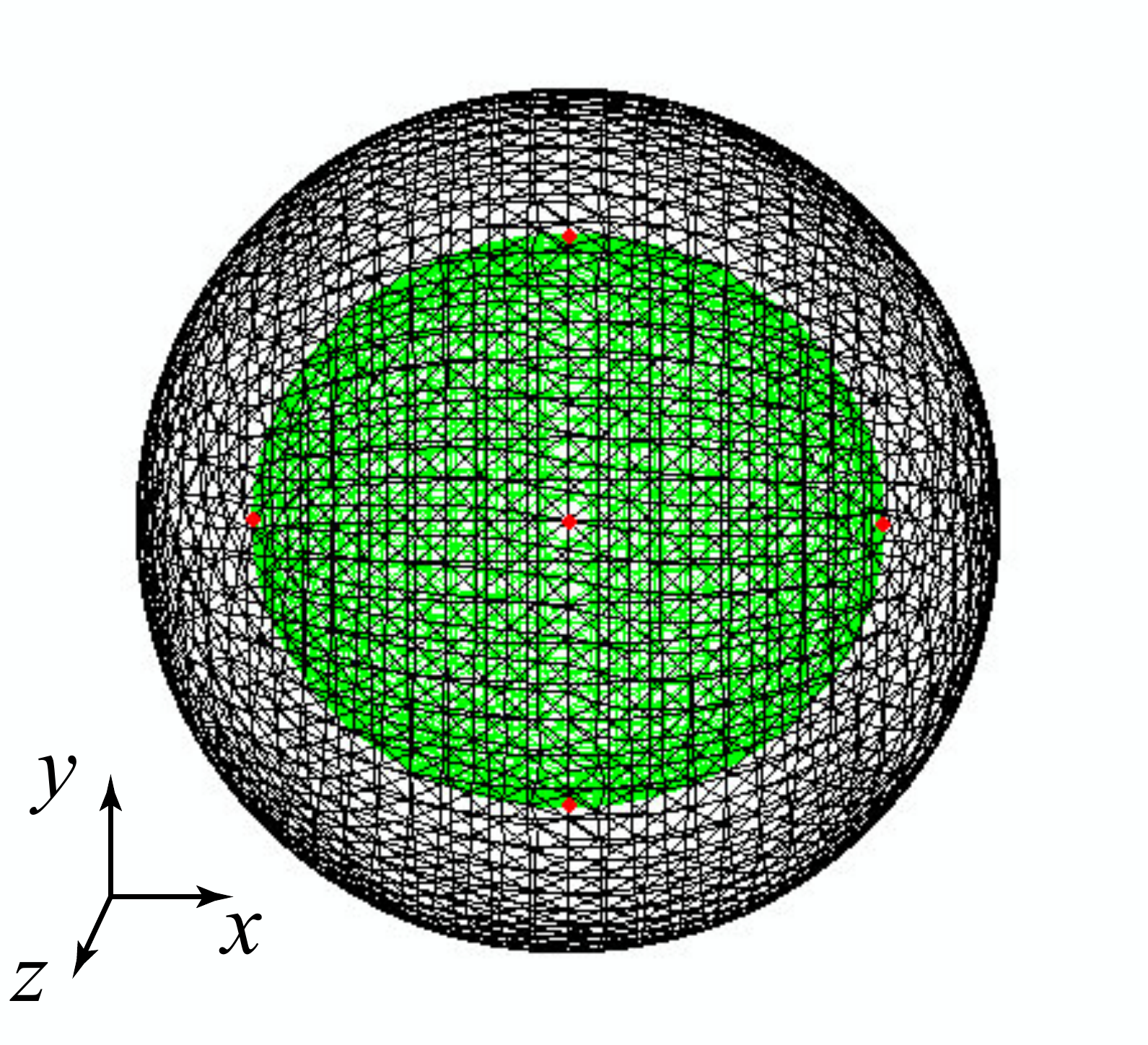}
b)\includegraphics[width=0.22\textwidth]{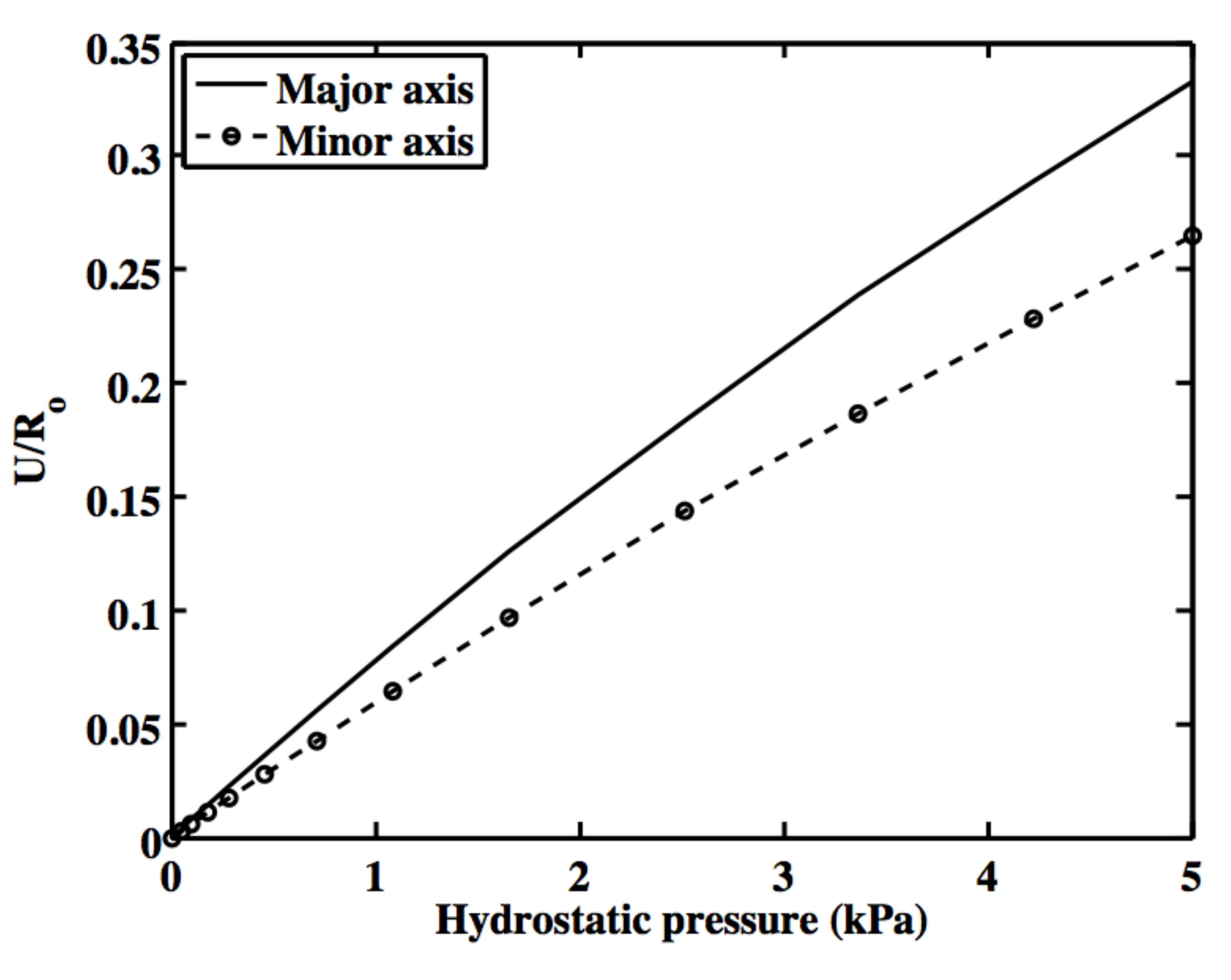}
}
\caption{
{\small In Abaqus\textsuperscript{\textregistered}, \emph{(a)} a sphere (black) with the {linear} elastic constants of paratellurite (transversely isotropic Tellurium dioxide TeO$_2$; see, for example, \cite{RoD84}) deforms into an ellipsoid (green) under hydrostatic pressure loading.  {Here the applied pressure is increased beyond the small strain regime to produce a ``visible'' deformation for illustrative purposes, but the ellipsoidal shape emerges immediately upon application of the pressure loading in the small deformation regime, as exemplified in  \emph{(b)}, \textcolor{black}{which displays the non-dimensionalized displacements of the end points (red dots on the major $x$-axis and the minor $y$- and $z$-axes) with respect to the hydrostatic stress (in kPa)}.}}}
\label{fig-TeO2-pressure}
\end{figure}

First, we put a sphere made of (transversely iso\-tropic) Tellurium Dioxide TeO$_2$ under hydrostatic loading. This material is linearly elastic and possesses the 422 tetragonal symmetry, with elastic stiffness constants $c_{11} = 5.59$, $c_{12} = 5.13$, $c_{13}=2.17$, $c_{33}=10.56$, $c_{44}=2.67$, $c_{66}= 6.62$ ($10^{10}$ N/m$^{2}$) and mass density $\rho=6020$ kg/m$^3$.
It is considered to be ``strongly anisotropic'' \cite{RoDi00}.
Figure \ref{fig-TeO2-tension} in the Introduction illustrates the behaviour for hydrostatic tension, while here in Figure \ref{fig-TeO2-pressure} the sphere is subject to hydrostatic pressure {$(p>0)$}.
As was the case for a hydrostatic tensile loading (Figure \ref{fig-TeO2-tension}), the sphere deforms into an ellipsoid under a hydrostatic pressure loading, as indeed it should. Again, in this simulation we pushed the deformation beyond the limit of validity of the linear theory for illustrative purposes, but the change of shape {from a sphere to an ellipsoid} was found to take place as soon as the deformation started, as attested by the plot of {\color{black}dimensionless displacements} of six points on the surface of the sphere shown in Figure \ref{fig-TeO2-pressure}(b). The six points on the surface of the sphere are highlighted in Figure \ref{fig-TeO2-pressure}(a) and they represent the end-points of the major axis and minor axes of the deformed ellipsoid.

\begin{figure}[!h]
\centerline{
a)\includegraphics[width=0.4\textwidth]{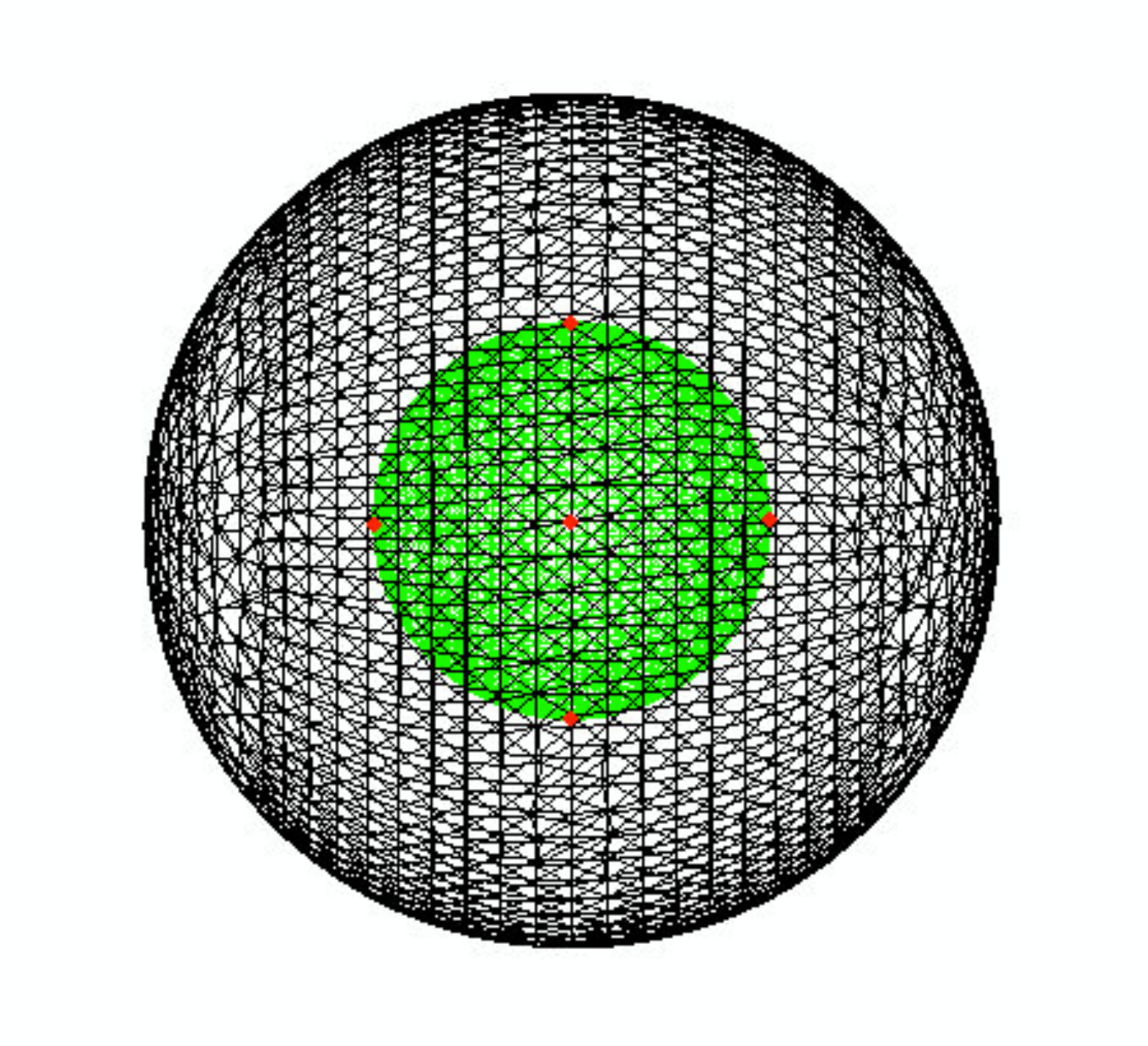}
b)\includegraphics[width=0.45  \textwidth]{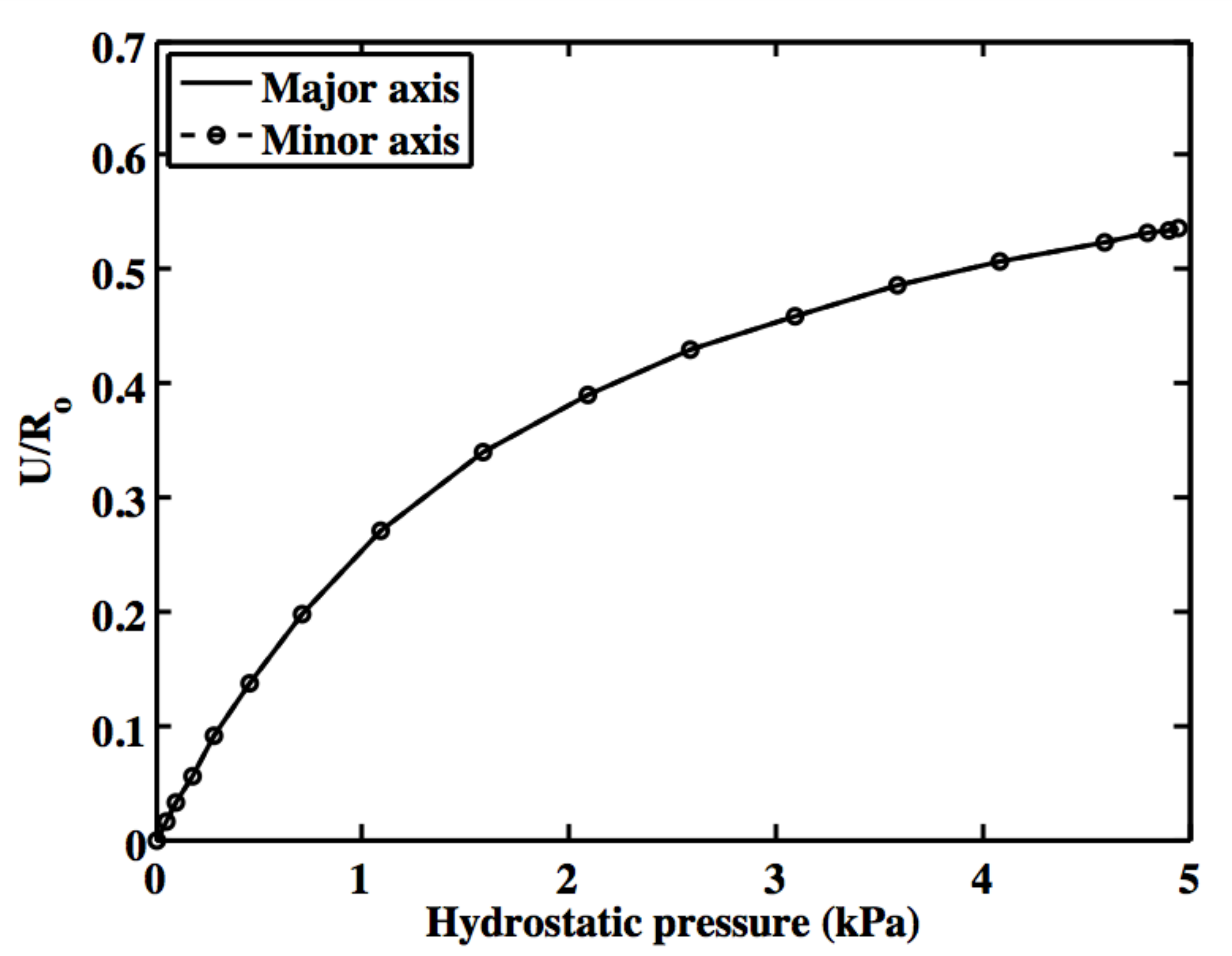}
}
\caption{
{\small In Abaqus\textsuperscript{\textregistered}, \emph{(a)} a sphere (black) modelled by the ``Holzapfel--Gasser--Ogden'' model deforms into a smaller sphere (green) under  hydrostatic pressure. This material is often used to model nonlinear soft tissues reinforced with one or two families of stiff fibres.
In the simulation, the loading produces a large deformation and the spherical shape remains, as exemplified in \emph{(b)}, which displays the non-dimensionalized displacement of the minor and major end-points with respect to the hydrostatic stress (in kPa).}}
\label{fig-holzapfel-pressure}
\end{figure}

Next we used a sphere made of Holzapfel--Ogden--Gasser material.
This is one of the most commonly used  hyperelastic anisotropic models, due to its excellent ability to capture the behaviour of arterial and other soft tissues. Based on structural observations, it assumes that for artery walls, there are two families of mechanically equivalent stiff collagen fibres embedded in a softer isotropic matrix.
An extension of the model (Gasser et al. \cite{holz2}) can also account for distributed collagen fibre orientations.
For our experiments we picked the following material and structural constants: $\kappa = 0$ (no dispersion), $\mu = 50$ kPa, $k_1 = 1$ MPa, $k_2 = 100$, $\Theta = 50^\circ$ \textcolor{black}{(see \cite{holz2} for details)}.
Figure \ref{fig-holzapfel-tension} shows the result for a sphere under hydrostatic tension, while Figure  \ref{fig-holzapfel-pressure} shows the corresponding result for hydrostatic pressure. In both cases, the sphere deforms into another sphere. To dispel any doubt about this phenomenon, we also display the displacement magnitudes of some representative points on the surface, and find that they all agree.

As we saw earlier, this clearly unphysical behaviour is due to the additive decomposition of equation \eqref{assumption2}, at least in the infinitesimal regime. However, when we increased the hydrostatic {\color{black}\emph{tension}} enough to produce a large deformation we also obtained a sphere, even though our proof does not carry over to the nonlinear regime, as we show in the following section.
For the case of hydrostatic {\color{black}\emph{pressure}} the situation is different since the fibre families in the Holzapfel--Gasser--Ogden model are assumed not to support compression and then the material response is isotropic;  it is then  expected that a sphere will deform into a smaller sphere.

\textcolor{black}{All finite element simulations in this study were initially performed using a low density mesh consisting of 35252 four noded tetrahedron elements with linear interpolation functions. A convergence analysis was performed using a high density mesh consisting of 209023 four noded tetrahedron elements with linear interpolation functions. Computed nodal deformations on the surface of the sphere for the high density mesh differed from those computed for the low density mesh by less than 0.0016\%, demonstrating a highly converged solution. Additionally, simulations revealed that the use of ten noded tetrahedron elements with quadratic interpolation functions resulted in a change of computed results of less than 0.0012\%.}

\section{Hydrostatic stress versus pure dilatation}
\label{Hydrostatic stress}


Here we show that a sphere made of a compressible nonlinear anisotropic material deforms into an ellipsoid under \color{black}a large \color{black} hydrostatic stress, whether or not its strain-energy function is decoupled.

For ease of illustration we consider a model in which the strain-energy function depends only on the invariants $I_1$, $I_3$, $I_4$, $I_6$ corresponding to a simplified model for an isotropic matrix with two embedded families of fibres.
This model includes the Holzapfel--Gasser--Ogden material as a special case.
Thus, $W=W(I_1,I_3,I_4,I_6)$, and the Cauchy stress $\boldsymbol{\sigma}$ is given by
\begin{equation} \label{sigma-2}
J\boldsymbol{\sigma}=2W_1\mathbf{B}+2I_3W_3\mathbf{I}+2W_4\vec{m}\otimes\vec{m}+2W_6\vec{m}'\otimes\vec{m}'.
\end{equation}
For a pure (and uniform) hydrostatic stress $\boldsymbol{\sigma}=\sigma\mathbf{I}$, define $\bar{\sigma}=J\sigma-2I_3W_3$.  Then the above becomes
\begin{equation}
\bar{\sigma}\mathbf{I} = 2W_1\mathbf{B}+2W_4\vec{m}\otimes\vec{m}+2W_6\vec{m}'\otimes\vec{m}'.
\end{equation}

We suppose the initial fibre directions $\vec{M}$ and $\vec{M}'$ lie in the plane defined by $\vec{e}_1$ and $\vec{e}_2$, \textcolor{black}{contain an angle of $2\Theta$} and have components
\begin{equation} \label{theta}
\vec{M}=(\cos\Theta,\sin\Theta,0),\quad \vec{M}'=(\cos\Theta,-\sin\Theta,0).
\end{equation}
We also assume that the two families of fibres have the same mechanical properties.  Then, by symmetry we can consider the deformation resulting from the hydrostatic stress to
be a pure homogeneous strain with principal directions $\vec{e}_1,\vec{e}_2,\vec{e}_3$ and corresponding stretches $\lambda_1,\lambda_2,\lambda_3$.  The deformed fibre directions are then
\begin{equation}
\left.\begin{array}{ll}
\vec{m}=(\lambda_1\cos\Theta,\lambda_2\sin\Theta,0),\\
 \vec{m}'=(\lambda_1\cos\Theta,-\lambda_2\sin\Theta,0),
 \end{array}\right.
\end{equation}
and the invariants $I_4$ and $I_6$ are given by
\begin{equation}
I_4=I_6=\vec{m}\cdot\vec{m}=\vec{m}'\cdot\vec{m}'=\lambda_1^2\cos^2\Theta+\lambda_2^2\sin^2\Theta.\label{I4I6theta}
\end{equation}

Since the two families of fibres have the same properties then, \emph{for the considered deformation}, $W_4=W_6$ and hence
\begin{equation} \label{equal-fibers}
2W_1\mathbf{B}+2W_4(\vec{m}\otimes\vec{m}+\vec{m}'\otimes\vec{m}')=\bar{\sigma}\mathbf{I}.
\end{equation}
It follows that
\begin{align}
& 2W_1\mathbf{B}(\vec{m}+\vec{m}')=[\bar{\sigma}-2W_4(I_4+\vec{m}\cdot\vec{m}')](\vec{m}+\vec{m}'),\\[0.5ex]
& 2W_1\mathbf{B}(\vec{m}-\vec{m}')=[\bar{\sigma}-2W_4(I_4-\vec{m}\cdot\vec{m}')](\vec{m}-\vec{m}'),
\end{align}
i.e. $\vec{m}+\vec{m}'$ and $\vec{m}-\vec{m}'$, which are along the bisectors of the fibres in the deformed configuration, are eigenvectors of $\mathbf{B}$.  Also $\vec{m}\times\vec{m}'$, which is along the normal to the plane of the fibres, is an eigenvector.  Let them correspond to the eigenvalues $\lambda_1^2,\lambda_2^2,\lambda_3^2$, respectively.  Then
\begin{eqnarray}
\lambda_1^2&=&\displaystyle\frac{\bar{\sigma}-2W_4(I_4+\vec{m}\cdot\vec{m}')}{2W_1},\notag\\
[1ex]
\lambda_2^2&=&\displaystyle\frac{\bar{\sigma}-2W_4(I_4-\vec{m}\cdot\vec{m}')}{2W_1},\notag\\
[1ex]
 \lambda_3^2&=&\displaystyle\frac{\bar{\sigma}}{2W_1},
\end{eqnarray}
and hence
\begin{align}
& \lambda_1^2-\lambda_2^2 =-2W_4(\vec{m}\cdot\vec{m}')/W_1,\nonumber \\[0.5ex]
& \lambda_1^2-\lambda_3^2 = -W_4(I_4+\vec{m}\cdot\vec{m}')/W_1,\nonumber \\[0.5ex]
& \lambda_2^2-\lambda_3^2 = -W_4(I_4-\vec{m}\cdot\vec{m}')/W_1,
\label{B11}
\end{align}
and we also have $\vec{m}\cdot\vec{m}'=\lambda_1^2\cos^2\Theta-\lambda_2^2\sin^2\Theta$.

As a consequence, for a fibre-reinforced material with $W_4\neq 0$, we note that $\lambda_1=\lambda_2$ if and only if $\vec{m}\cdot\vec{m}'=0$, in which case we must have $\cos2\Theta=0$, i.e. the fibres are initially orthogonal (a special case).  But also, $\lambda_1=\lambda_2\neq\lambda_3$ except when $W_4=0$ (which takes us back to isotropy).  A special case also arises if $\vec{m}\cdot\vec{m}'\neq 0$ and $I_4=\pm \vec{m}\cdot\vec{m}'$, but this corresponds to the two fibre families being aligned ($\Theta=0$ or $\Theta=\pi/2$).
On the other hand, \emph{if} the fibres are initially orthogonal then we obtain
\begin{equation}
(\lambda_1^2-\lambda_2^2)(W_1+W_4)=0,
\end{equation}
and we deduce that $\lambda_1=\lambda_2 \neq \lambda_3$.
With appropriate slight changes the above analysis applies also for a single fibre family.

Now consider the energy function, again denoted $W^*$, to be expressed in terms of the deviatoric invariants $I_1^*=I_3^{-1/3} I_1$, $I_4^*=I_3^{-1/3}I_4$, $I_6^*=I_3^{-1/3}I_6$, and $J$. Then, we have \cite{murph12}
\begin{multline}
J\boldsymbol{\sigma}= JW^*_J + 2W_1^*\left(\mathbf{B}^*-\tfrac{1}{3}I_1^*\mathbf{I}\right) \\
+2W_4^*\left(\vec{m}^*\otimes\vec{m}^*-\tfrac{1}{3}I_4^*\mathbf{I} \right)\\
+2W_6^*\left({\vec{m}'}^*\otimes{\vec{m}'}^*-\tfrac{1}{3}I_6^*\mathbf{I}\right), \label{B13}
\end{multline}
where  $\mathbf{B}^* = J^{-2/3}\mathbf{B}$, $\vec{m}^* =  J^{-1/3}\vec{m}$,  ${\vec{m}'}^*=  J^{-1/3}\vec{m}' $.
If the strain energy is decoupled into a deviatoric and a volumetric part as
\begin{equation} \label{B14}
W^*(I_1^*, I_4^*, I_6^*;J) = f(J) + \mathcal W(I_1^*, I_4^*, I_6^*),
\end{equation}
then the result of our proof above is unaffected, because \eqref{B11} is simply replaced by
\begin{align}
& \lambda_1^2-\lambda_2^2 = -2 \mathcal W_4(\vec{m}\cdot{\vec{m}'}^*)/\mathcal W_1,\nonumber \\
& \lambda_1^2-\lambda_3^2 = -\mathcal W_4(I^*_4+\vec{m}^*\cdot{\vec{m}'}^*/\mathcal W_1,\nonumber \\
& \lambda_2^2-\lambda_3^2 = -\mathcal W_4(I^*_4-\vec{m}\cdot\vec{m}')/\mathcal W_1,
\end{align}
and the same deductions follow.

Now take the trace of both sides of equation \eqref{B13} for the case when the strain energy is decoupled according to  \eqref{B14}.  This gives
\begin{equation} \label{B16}
\tfrac{1}{3}\tr\boldsymbol{\sigma} = f'(J).
\end{equation}
\textcolor{black}{Thus, as pointed out in \cite{sans08}, when there is a volumetric/deviatoric split, the deviatoric stress changes only the shape and not the volume, in contrast to the situation where there is no such split.}

N\'i Annaidh et al. \cite{murph12} argued that the mean Cauchy stress on the left-hand side of this equation should cause more than the simple volume change that the right-hand side would seem to indicate. However, even though the mean Cauchy stress depends only on $J$ here, the response of the material is not in general purely dilatational, as we now show through an example.

For simplicity we consider a single family of fibres, so that
\begin{equation}
J\boldsymbol{\sigma}=2 \mathcal{W}_1(\mathbf{B}^*-\tfrac{1}{3}I_1^*\mathbf{I})+2 \mathcal{W}_4(\vec{m}^*\otimes\vec{m}^*-\tfrac{1}{3}I_4^*\mathbf{I})
+ Jf'(J).
\end{equation}
Suppose we consider a uniaxial stress $\sigma_1$ along the $\vec{m}$ direction.  Then this equation gives (with $\lambda_2=\lambda_3$ by symmetry, $I_4=\lambda_1^2$ and $\sigma_2=\sigma_3=0$)
\begin{align}
& J\sigma_1=\tfrac{4}{3}\mathcal{W}_1I_3^{-1/3}(\lambda_1^2-\lambda_2^2)+\tfrac{4}{3} \mathcal{W}_4I_3^{-1/3}\lambda_1^2+J f'(J), \nonumber \\[8pt]
&
0=-\tfrac{2}{3} \mathcal{W}_1I_3^{-1/3}(\lambda_1^2-\lambda_2^2)-\tfrac{2}{3} \mathcal{W}_4I_3^{-1/3}\lambda_1^2+J f'(J).
\end{align}
By subtraction, and use of \eqref{B16}, we obtain the two equalities
\begin{equation}
J\sigma_1 = 2 \mathcal{W}_1I_3^{-1/3}(\lambda_1^2-\lambda_2^2) + 2 \mathcal{W}_4I_3^{-1/3}\lambda_1^2 = 3 J f'(J).
\end{equation}
Since $\lambda_2=\lambda_3$ and hence $I_3=\lambda_1^2\lambda_2^4$, the latter equation determines $\lambda_2$ in terms of $\lambda_1$, at least in principle.  Thus, $\sigma_1$ is a function of $\lambda_1$ and $\lambda_2$, but in general  $\lambda_1\neq\lambda_2$.
Equation \eqref{B16} shows that the dilatation is determined by the hydrostatic part of the stress, but that does not imply that the deformation is a pure dilatation or that the uniaxial stress depends only on the dilatation, because of the connection between $\lambda_1$ and $\lambda_2$.

We conclude this section by studying the effects of an assumption which is often made when modelling biological soft tissues: that \emph{collagen fibres don't withstand compression}.
Mathematically this is translated by letting the material have an effective fibre contribution to its strain-energy function (as in equation~\eqref{sigma-2}) when the stretch is greater than unity in the direction of the fibres, and by taking that contribution to be zero when the stretch is smaller than unity.
Because this stretch is $\sqrt{I_4}$ in the direction of $\vec{M}$ and $\sqrt{I_6}$ in the direction of $\vec{M}'$, it means that here equation~\eqref{sigma-2} is in place when $I_4>1$ and $I_6>1$.
When $I_4 \le 1$ (compression in the direction of $\vec M$), equation~\eqref{sigma-2} is reduced by one term by effectively taking $W_4 \equiv 0$ and similarly when $I_6 \le 1$ (compression in the direction of $\vec M'$), in which case we take $W_6 \equiv 0$.

For a material with two families of parallel fibres with equal properties, suppose again that the initial fibre directions $\vec{M}$ and $\vec{M}'$ are as in equation~\eqref{theta}, and that the deformation resulting from a hydrostatic stress is a pure homogeneous strain with corresponding stretches $\lambda_1,\lambda_2,\lambda_3$, and $I_4=I_6 >1$. Now we write the components of \eqref{equal-fibers} as
\begin{eqnarray}
\bar{\sigma}&=&2W_1\lambda_3^2,\notag\\[1ex]
\bar{\sigma}&=&2W_1\lambda_1^2+4W_4\lambda_1^2\cos^2\Theta,\notag\\[1ex]
\bar{\sigma}&=&2W_1\lambda_2^2+4W_4\lambda_2^2\sin^2\Theta.\label{equations}
\end{eqnarray}

Assume first that the deformation determines a \emph{volume compression}, as in Figures \ref{fig-TeO2-pressure} and \ref{fig-holzapfel-pressure}, or leaves the volume unchanged: $J\leq 1$.
Then, compare the invariants $I_4=I_6$ to 1.
If $I_4>1$, then $W_4 \ne 0$ in \eqref{equations}, which we can solve to find, using \eqref{equations} and the connection $J =\lambda_1\lambda_2\lambda_3$, that
\begin{align} \label{lambdas}
& \lambda_1=J^{1/3}\left[\frac{W_1(W_1+2W_4\sin^2\Theta)}{(W_1+2W_4\cos^2\Theta)^2}\right]^\frac{1}{6},\notag \\[8pt]
&  \lambda_2=J^{1/3}\left[\frac{W_1(W_1+2W_4\cos^2\Theta)}{(W_1+2W_4\sin^2\Theta)^2}\right]^\frac{1}{6},\notag \\[8pt]
& \lambda_3=J^{1/3}\left[\frac{W_1^2+2W_1W_4+W_4^2\sin^22\Theta}{W_1^2}\right]^\frac{1}{6}.
 \end{align}
By substituting for $\lambda_1$ and $\lambda_2$ in the expression for $I_4$ in  \eqref{I4I6theta} we then obtain
\begin{equation}\label{in}
I_4=J^{2/3}g(\Theta)>1,
\end{equation}
where
\begin{eqnarray}
g(\Theta)&=&\left[\frac{W_1(W_1+2W_4\sin^2\Theta)}{(W_1+2W_4\cos^2\Theta)^2}\right]^\frac{
1}{3}\cos^2\Theta \notag\\
& &+\left[\frac{W_1(W_1+2W_4\cos^2\Theta)}{(W_1+2W_4\sin^2\Theta)^2}\right]^\frac{1}{3}\sin^2\Theta.
\end{eqnarray}
However, we can easily compute the extrema of $g$ and establish that
\begin{equation}
0<\left[\frac{W_1^2}{(W_1+2W_4)^2}\right]^\frac{1}{3}\leq g(\Theta)\leq \left[\frac{W_1}{W_1+W_4}\right]^\frac{1}{3}<1,
\end{equation}
for all $\Theta$. Then inequality \eqref{in}  leads necessarily to $J>1$, contradicting the assumption of volume compression.
It follows that $I_4=I_6\leq1$ so that $W_4 \equiv 0$ in \eqref{equations},  giving $\lambda_1=\lambda_2=\lambda_3=J^{1/3}$: \emph{in compression therefore, a sphere changes into a \color{black} smaller \color{black} sphere under hydrostatic stress}.
This observation goes a long way to explain Figure \ref{fig-holzapfel-pressure} for the sphere of Holzapfel--Gasser--Ogden material in compression.

Assume next that the deformation determines a \emph{volume expansion}: $J>1$, as in Figures \ref{fig-TeO2-tension} and \ref{fig-holzapfel-tension}. Then, compare the invariants $I_4=I_6$ to 1. If $I_4 \le 1$, then $W_4 \equiv 0$ and by \eqref{equations}, all three stretches would be equal: $\lambda_1=\lambda_2=\lambda_3=J^{1/3}$ (recall that  $\lambda_1\lambda_2\lambda_3=J$).
But then, it would follow from \eqref{I4I6theta} that $I_4= \lambda_1^2 =J^{2/3}>1$, a contradiction.
We {\color{black}therefore} deduce that $I_4 = I_6 >1$ and thus we take $W_4=W_6\neq0$ in \eqref{equations}.
We then find, using \eqref{equations} and the identity $J =\lambda_1\lambda_2\lambda_3$, that
the expressions \eqref{lambdas} hold.
Therefore, for any value of $\Theta$ at least two of the principal stretches are different:
\emph{in expansion, a sphere changes into an ellipsoid under hydrostatic stress}.

The overall conclusion of this section is that \emph{in compressible nonlinear anisotropic hyperelasticity,  hydrostatic tension is not accompanied by a pure dilatation}, whether or not the strain-energy function is decoupled into a deviatoric part and a volumetric part; and under hydrostatic pressure the same conclusion follows unless the fibres do not support compression.
This is consistent with the work of Sansour \cite{sans08} who showed that for orthotropic materials, a purely spherical state of stress is accompanied by a change of shape.

The case of an incompressible material is covered by the first of the above arguments (for $J\leq 1$) and for $J=1$ we conclude that $\lambda_1=\lambda_2=\lambda_3=1$: \emph{for an incompressible material under hydrostatic stress, whether positive or negative, there is no deformation and a sphere {\color{black}remains undeformed}}.

However, a question remains about the way Abaqus\textsuperscript{\textregistered}  deals with compressible anisotropic hyperelasticity using the volumetric/deviatoric separation.  As we have shown for the infinitesimal theory, for a decoupled model under hydrostatic stress a sphere deforms into another sphere.  Abaqus\textsuperscript{\textregistered}  carries this result over to the finite deformation regime not only in compression, which can be explained by the fact that fibres do not support contraction, but also in expansion, contrary to theoretical predictions.


\section{Why a sphere deforms into a sphere in FE simulations of nonlinear anisotropic elasticity}
\label{why}


The Holzapfel--Gasser--Ogden strain-energy function\linebreak used for modelling arterial layers with two families of parallel collagen fibres, with the same mechanical properties, has the general form \cite{holz1}
\begin{equation}\label{strain energy}
W(I_1,I_3,I_4,I_6)=f(J)+\mathcal{W}(I_1^*,I_4^*,I_6^*),
\end{equation}
where $f(J)$ is the volumetric part and
\begin{equation}\label{strain}
 \mathcal{W}=\frac{\mu}{2}(I_1^*-3)+ \Psi^*_1 + \Psi^*_2,
\end{equation}
showing the respective contributions of an isotropic neo-Hookean matrix and of the reinforcing fibres.  For the version of this model including fibre dispersion \cite{holz2},
$\Psi^*_1$ and $ \Psi^*_2$ have the forms
\begin{align}\label{aniso}
& \Psi^*_1 = \dfrac{k_1}{2k_2}\left[\exp\left\{k_2\left[\kappa I_1^* + (1-3\kappa)I^*_4 -1\right]^2\right\} - 1\right], \notag \\
& \Psi^*_2 = \dfrac{k_1}{2k_2}\left[\exp\left\{k_2\left[\kappa I_1^* + (1-3\kappa)I^*_6 -1\right]^2\right\} - 1\right],
\end{align}
and $\mu$, $\kappa$, $k_1$, $k_2$ are material parameters.
This model is implemented in Abaqus\textsuperscript{\textregistered}.
It is also implemented in ADINA\textsuperscript{\textregistered} with $\kappa=0$, i.e. there is no dispersion in the fibre distribution.

What  such models have in common is that their anisotropic part enters into play only \emph{when the deviatoric stretch is greater than unity} in the fibre direction(s), that is when $\I_4>1$ and/or $\I_6>1$.
This subtle difference with the previous section (where anisotropy contributed to the strain energy when the \emph{actual fibre stretches} were greater than unity, that is when $I_4>1$ and/or $I_6>1$) has dramatic repercussions for the hydrostatic tensile loading of a sphere, as we now see.

Re-consider  the hydrostatic loading case examined in the previous section in which the deformation determines a volume expansion: $J>1$ with $I_4=I_6>1$ and hence $\W_4>0$.
The deviatoric invariants read
\begin{equation} \label{6.5}
\I_4 = \I_6 = {\lambda_1^*}^2\cos^2\Theta+{\lambda_2^*}^2\sin^2\Theta,
\end{equation}
where the deviatoric principal stretches $\lambda_i^* \equiv J^{-1/3}\lambda_i$, $i=1,2$, are found, similarly to \eqref{lambdas}, as
\begin{eqnarray} 
\lambda^*_1&=&\left[\dfrac{\W_1(\W_1+2\W_4\sin^2\Theta)}{(\W_1+2\W_4\cos^2\Theta)^2}\right]^\frac{1}{6},\notag\\
  \lambda^*_2&=&\left[\dfrac{\W_1(\W_1+2\W_4\cos^2\Theta)}{(\W_1+2\W_4\sin^2\Theta)^2}\right]^\frac{1}{6}.\label{6.6}
  \end{eqnarray}
Substituting \eqref{6.6} into \eqref{6.5}, we deduce that
\begin{multline}
I_4^* = I_6^* = \left[\frac{\mathcal{W}_1(\mathcal{W}_1+2\mathcal{W}_4\sin^2\Theta)}{(\mathcal{W}_1+2\mathcal{W}_4\cos^2\Theta)^2}\right]^\frac{1}{3}\cos^2\Theta \\
+ \left[\frac{\mathcal{W}_1(\mathcal{W}_1+2\mathcal{W}_4\cos^2\Theta)}{(\mathcal{W}_1+2\mathcal{W}_4\sin^2\Theta)^2} \right]^\frac{1}{3}\sin^2\Theta
 \leq \left[\frac{\mathcal{W}_1}{\mathcal{W}_1+\mathcal{W}_4} \right]^\frac{1}{3},\label{I4star-I6star}
\end{multline}
which is strictly less than 1 for all $\Theta$.
Hence \emph{the deviatoric stretches are always less than unity} in the fibre directions, and the anisotropic contribution to the strain energy function \eqref{strain energy} disappears.
Consequently, we obtain $\lambda_1=\lambda_2=\lambda_3=J^{1/3}$, which is why our Abaqus\textsuperscript{\textregistered} simulations show that a sphere subject to hydrostatic tension deforms into a sphere of greater radius instead of an ellipsoid. Although we have not tried this experiment with ADINA\textsuperscript{\textregistered}, it is clear that it will predict the same unphysical pure dilatation.

Finally, we note that when the deformation determines a volume compression or leaves the volume unchanged ($J\leq1$) then  $I_4^*=I_6^*=1$ since $I_4=I_6=J^{2/3}$ so that the anisotropic contribution in the strain-energy function \eqref{strain energy}--\eqref{strain}  also disappears.

{\color{black}It is important to remark here that while we have used the Holzapfel--Gasser--Ogden model for illustration, the above discussion applies to any model that is a function of $I_1^*$, $I_3$, $I_4^*$ and $I_6^*$ for which the anisotropic contribution is suppressed when $I_4^*\leq 1$ and $I_6^*\leq 1$, whether or not the strain-energy function is decoupled.  In particular, $\W$ may be replaced by $W$ in \eqref{I4star-I6star}.}




\section*{Appendix A:  Monoclinic elasticity }

\numberwithin{equation}{section}
\setcounter{equation}{1}

\renewcommand{\theequation}{A.\arabic{equation}}

Here we extend the results of Sections \ref{Inconsistency with transverse isotropy} and \ref{Inconsistency with orthotropy} to the case of monoclinic symmetry, for which, in the linear theory, there are 13 independent elastic constants.  For this purpose it suffices to consider an isotropic matrix material reinforced with two families of fibres, with fibre directions defined by $\vec{M}$ and $\vec{M}'$ in the reference configuration, the fibres being in general neither at right angles nor mechanically equivalent.  First we establish the general equations of compatibility between nonlinear and linear anisotropic elasticity.

With two fibre directions, the strain-energy function $W$ is a function of three isotropic strain invariants ($I_1, I_2, I_3$) and five anisotropic invariants ($I_4$, $I_5$, $I_6$, $I_7$, $I_8)$.  Thus,
\begin{equation}
W = W(I_1,I_2,I_3,I_4,I_5, I_6,I_7,I_8),
\end{equation}
where
 \begin{align}
& I_1 = \tr \, \mathbf{C}, \quad  I_2 = \frac{1}{2}\left[\left(\tr \, \vec  C\right)^2  -\tr\left(\mathbf{C}^2\right)\right],\notag
\\[1ex]
& I_3 = \det \mathbf{C},\quad  I_4 =  \vec{M} \cdot \mathbf{C}\vec{M}, \quad  I_5 = \vec {M}\cdot\mathbf{C}^2 \vec{M},\notag \\[1ex]
&  I_6 =  \vec{M}' \cdot \mathbf{C}\vec{M}', \quad  I_7 = \vec {M}'\cdot \mathbf{C}^2 \vec{M}',\quad  I_8 =  \vec{M} \cdot \mathbf{C}\vec{M}'.
\label{I1I8}
\end{align}
Note that, to simplify the ensuing analysis, we are using the invariant $I_8$ as defined above, rather than one of its strictly invariant forms $I_8\vec{M}\cdot\vec{M}'$ or $I_8^2$, which do not depend on the \emph{sense} of either $\vec{M}$ or $\vec{M}'$.

For the expression for the Cauchy stress $\boldsymbol{\sigma}$ in the incompressible case we refer to Merodio and Ogden \cite{MeOg06}.  Here we use its compressible counterpart
\begin{eqnarray}
\label{cauch}
J\boldsymbol{\sigma}&=&2W_1\mathbf{B}+2W_2(I_1\mathbf{B}-\mathbf{B}^2)+2I_3W_3\mathbf{I}
\nonumber \\[0.5ex]
& +&2W_4\vec{m}\otimes\vec{m} +2W_5(\mathbf{B}\vec{m}\otimes\vec{m}+\vec{m}\otimes\mathbf{B}\vec{m})\nonumber\\[0.5ex]
&+&2W_6\vec{m}'\otimes\vec{m}' \nonumber  +2W_7(\mathbf{B}\vec{m}'\otimes\vec{m}'+\vec{m}'\otimes\mathbf{B}\vec{m}')\nonumber\\[0.5ex]
&+&W_8(\vec{m}\otimes\vec{m}'+\vec{m}'\otimes\vec{m}).
\end{eqnarray}

In the reference configuration the invariants take the values
\begin{equation}
I_1=I_2=3,\ \
I_3=I_4=I_5=I_6=I_7=1, \ \
 I_8= \vec{M}\cdot\vec{M}',\label{ref-mono}
\end{equation}

Assuming that the reference configuration is stress free, it follows from \eqref{cauch}, when evaluated in the reference configuration, that the conditions
\begin{eqnarray}
&&W_1+2W_2+W_3=0,\notag\\[1ex]
 &&W_4+2W_5=0, \quad W_6+2W_7=0,\quad W_8=0
\end{eqnarray}
must hold there.

Now let $\mathbf{e}$ be the infinitesimal strain tensor and let $e=\tr\mathbf{e}$.   Then, to the first order in $\mathbf{e}$, we obtain
\begin{align}
& I_1=3+2e, \qquad I_2=3+4e,\qquad I_3=1+2e,\nonumber \\
&  I_4=1+2\emm,\qquad  I_5=1+4\emm,\nonumber \\
& I_6=1+2\empmp,\qquad  I_7=1+4\empmp,\nonumber \\
&I_8=\vec{M}\cdot\vec{M}'+2\emmp,
 \end{align}
and $\mathbf{B}=\mathbf{I}+2\mathbf{e}$.  Then, using the restrictions \eqref{restrict} and linearizing \eqref{cauch} in $\mathbf{e}$, we obtain, after a lengthy but straightforward process,
\begin{align}
\nonumber
&\boldsymbol{\sigma}=
\alpha\mathbf{e}+\big[(\beta-\alpha)e+\gamma (\emm)+\gamma' (\empmp)\\[0.5ex]
\nonumber
&+\zeta\emmp\big]\mathbf{I}+4W_5(\vec{M}\otimes\mathbf{e}\vec{M}+\mathbf{e}\vec{M}\otimes\vec{M})\\[0.5ex]
\nonumber
&+4W_7(\vec{M}'\otimes\mathbf{e}\vec{M}'+\mathbf{e}\vec{M}'\otimes\vec{M}')\\[0.5ex]
\nonumber
&+(\gamma e+\delta \emm +\epsilon \empmp+\eta\emmp)\vec{M}\otimes\vec{M}\\[0.5ex]
\nonumber
& + (\gamma' e +\epsilon \emm +\delta' \empmp+\eta'\emmp)\vec{M}'\otimes\vec{M}'\\[0.5ex]
\nonumber
&+\tfrac{1}{2}(\zeta e+\eta\emm+\eta'\empmp+4W_{88}\emmp)\\[0.5ex]
\nonumber
& \times (\vec{M}\otimes\vec{M}'+\vec{M}'\otimes\vec{M}),
\end{align}
where we have introduced the notations
\begin{align}
\nonumber
& \alpha=4(W_1+W_2),\\[0.5ex]
\nonumber
& \beta=4(W_{11}+4W_{12}+4W_{22}+2W_{13}+4W_{23}+W_{33}),
\nonumber \\[0.5ex]
& \gamma=4(W_{14}+2W_{24}+W_{34}+2W_{15}+4W_{25}+2W_{35}),
\nonumber \\[0.5ex]
&\gamma'=4(W_{16}+2W_{26}+W_{36}+2W_{17}+4W_{27}+2W_{37}),
\nonumber \\[0.5ex]
& \delta=4(W_{44}+4W_{45}+4W_{55}),\notag\\[0.5ex]
\nonumber
& \delta'=4(W_{66}+4W_{67}+4W_{77}),
\nonumber \\[0.5ex]
& \epsilon=4(W_{46}+2W_{47}+ 2W_{56}+4W_{57}),\notag\\[0.5ex]
& \zeta=4(W_{18}+2W_{28}+W_{38}),
\nonumber \\[0.5ex]
& \eta=4(W_{48}+2W_{58}), \quad \eta'=4(W_{68}+2W_{78})\label{greek-constants}
\end{align}
for the combinations of derivatives of $W$ evaluated in the reference configuration.  These constants, together with $W_5$, $W_7$ and $W_{88}$ constitute the 13 independent elastic constants of monoclinic symmetry.

By comparing with the general expression for the Cauchy stress in linear anisotropic elasticity in terms of the Voigt notation we obtain 21 expressions for elastic constants, only 13 of which are independent.
These are summarized as
\begin{eqnarray}
c_{ii}&=&\beta+2(\gamma +4W_5)M_i^2+2(\gamma'+4W_7){M_i'}^2+2\zeta M_iM_i'\notag\\[0.5ex]
&+&\delta M_i^4+ 2\eta M_i^3M_i'+(2\epsilon +4W_{88})M_i^2{M_i'}^2\notag\\[0.5ex]
&+&2\eta'M_i{M_i'}^3+\delta' {M_i'}^4,
\end{eqnarray}
\begin{eqnarray}
c_{ij}&=&\beta-\alpha+\gamma (M_i^2+M_j^2)+\gamma' ({M_i'}^2+{M_j'}^2)\notag\\[0.5ex]
&+&\zeta(M_iM_i'+M_jM_j')+\delta M_i^2M_j^2+\delta' {M_i'}^2{M_j'}^2\notag \\[0.5ex]
&+&\epsilon (M_i^2{M_j'}^2+ {M_i'}^2M_j^2) +4W_{88}M_iM_jM_i'M_j'\notag\\[0.5ex]
&+&(\eta M_iM_j+\eta'M_i'M_j')(M_iM_j'+M_i'M_j)
\end{eqnarray}
for $i,j\in\{1,2,3\},\,i\neq j$,
\begin{eqnarray}
c_{14}&=&(\gamma+\delta M_1^2 +\epsilon {M_1'}^2+\eta M_1M_1')M_2M_3\notag\\[0.5ex]
&+&
(\gamma'+\epsilon M_1^2 +\delta'{M_1'}^2+\eta' M_1M_1')M'_2M'_3\notag\\[0.5ex]
&+&\tfrac{1}{2}(\zeta +\eta M_1^2+\eta' {M_1'}^2+4W_{88}M_1M_1')\notag\\[0.5ex]
&\times&(M_2M_3'+M_2'M_3),
\end{eqnarray}
\begin{eqnarray}
c_{15}&=&(\gamma+4W_5+\delta M_1^2 +\epsilon {M_1'}^2+\eta M_1M_1')M_1M_3\notag\\[0.5ex]
&+&
(\gamma'+4W_7+\epsilon M_1^2 +\delta'{M_1'}^2+\eta' M_1M_1')M'_1M'_3\notag\\[0.5ex]
&+&\tfrac{1}{2}(\zeta +\eta M_1^2+\eta' {M_1'}^2+4W_{88}M_1M_1')\notag\\[0.5ex]
&\times&(M_1M_3'+M_1'M_3),
\end{eqnarray}
\begin{eqnarray}
c_{16}&=&(\gamma+4W_5+\delta M_1^2 +\epsilon {M_1'}^2+\eta M_1M_1')M_1M_2\notag\\[0.5ex]
&+&
(\gamma'+4W_7+\epsilon M_1^2 +\delta'{M_1'}^2+\eta' M_1M_1')M'_1M'_2\notag\\[0.5ex]
&+&\tfrac{1}{2}(\zeta +\eta M_1^2+\eta' {M_1'}^2+4W_{88}M_1M_1')\notag\\[0.5ex]
&\times&(M_1M_2'+M_1'M_2).
\end{eqnarray}
Note, in particular, that $W_5$ and $W_7$ do not appear in $c_{14}$.  This is because the index 4 corresponds to the pair of indices 23, which are different from the first index 1 in this case.  The constants $c_{24},c_{25},c_{26}$ and $c_{34},c_{35},c_{36}$ follow the same pattern, with the index 1 in the bracketed terms replaced by 2 and 3 respectively. Then,
$c_{25}$ and $c_{36}$ do not contain $W_5$ and $W_7$.

We also have
\begin{eqnarray}
c_{44}&=&\tfrac{1}{2}\alpha+2W_5(M_2^2+M_3^2)+2W_7({M_2'}^2+{M_3'}^2)\notag\\[0.5ex]
&+&\delta M_2^2M_3^2+2\epsilon M_2M_3M_2'M_3'+\delta' {M_2'}^2{M_3'}^2\notag\\[0.5ex]
&+&W_{88}(M_2M_3'+M_2'M_3)^2\notag\\[0.5ex]
&+&(\eta M_2M_3+\eta'M_2'M_3')(M_2M_3'+M_2'M_3).
\end{eqnarray}
Then, $c_{55}$ and $c_{66}$ are obtained by replacing the index 2 by 1 and the index 3 by 1, respectively.

Finally, we have
\begin{eqnarray}
c_{45}&=&(2W_5+\delta M_3^2+\eta M_3M_3')M_1M_2\notag\\[0.5ex]
&+&(2W_7+\delta {M_3'}^2+\eta' M_3M_3')M'_1M'_2\notag\\[0.5ex]
&+&\tfrac{1}{2}(\eta M_3^2+\eta'{M_3'}^2+2\epsilon M_3M_3')(M_1M_2'+M_1'M_2)\notag\\[0.5ex]
&+&W_{88}(M_1M_3'+M_1'M_3)(M_2M_3'+M_2'M_3),
\end{eqnarray}
and $c_{46}$ and $c_{56}$ are obtained simply by re-ordering the indices appropriately.

It is now convenient to let the two fibre directions in the reference configuration define the $(x_1,x_2)$ coordinate plane, so that $M_3=M_3'=0$ and the 21 constants reduce to the 13 appropriate for monoclinic symmetry, with $c_{14}=c_{24}=c_{34}=c_{15}=c_{25}=c_{35}=c_{46}=c_{56}=0$.

We now turn to the formulation of the strain-energy function based on the invariants $\I_1,\I_2, I_3, \I_4,\I_5,\I_6,\I_7$ defined in Section \ref{Inconsistency with orthotropy}, supplemented by their counterpart for $I_8$, namely
\begin{equation}
\I_8=J^{-2/3}I_8.
\end{equation}
Thus, $W^*=W^*(\I_1,\I_2, I_3, \I_4,\I_5,\I_6, \I_7,\I_8)$ and it easy to show that the conditions \eqref{restrict} holding in the reference configuration become
\begin{equation}\label{restrict-star}
W_3^*=0,\quad W^*_4+2W^*_5=0, \quad W^*_6+2W^*_7=0, \quad W^*_8=0.
\end{equation}
It also follows that
\begin{eqnarray}
W^*_{34}+2W^*_{35}&=&\tfrac{1}{12}(3\gamma+8W_5+\delta+\epsilon+\eta \vec{M}\cdot\vec{M}')\notag\\[0.5ex]
W^*_{36}+2W^*_{37}&=&\tfrac{1}{12}(3\gamma'+8W_7+\delta'+\epsilon+\eta' \vec{M}\cdot\vec{M}')\notag\\[0.5ex]
W^*_{38}&=&\tfrac{1}{12}(3\zeta +\eta+\eta'+4W^*_{88}\vec{M}\cdot\vec{M}'),\label{Wstarconditions}
\end{eqnarray}
and
\begin{eqnarray}
W^*_{33}&=&\tfrac{1}{4}\beta+\tfrac{1}{6}(\gamma+\gamma')+\tfrac{2}{9}(W_5+W_7)+\tfrac{1}{36}(\delta+\delta'+2\epsilon)\notag\\[0.5ex]
&+&\tfrac{1}{18}(3\zeta+\eta+\eta')\vec{M}\cdot\vec{M}'\notag\\[0.5ex]
&+&\tfrac{1}{9}W^*_{88}(\vec{M}\cdot\vec{M}')^2,
\end{eqnarray}
while the terms in \eqref{greek-constants} that do not involve a derivative with respect to $I_3$ are unaffected by the change $W\rightarrow W^*$. Note that $W^*_{13}$ and $W^*_{23}$ do not appear.

The terms in \eqref{Wstarconditions} can now be simply related to the Voigt constants.  Specifically, we obtain
\begin{align}
\tfrac{1}{12}(c_{11}+c_{12}&-c_{33}-c_{23})=(W^*_{34}+2W^*_{35})M_1^2\notag \\[0.5ex]
&+(W^*_{36}+2W^*_{37}){M_1'}^2+W^*_{38}M_1M_1',
\end{align}
\begin{align}
\tfrac{1}{12}(c_{22}+c_{12}&-c_{33}-c_{13})=(W^*_{34}+2W^*_{35})M_2^2
\nonumber\\
&+(W^*_{36}+2W^*_{37}){M_2'}^2+W^*_{38}M_2M_2',
\end{align}
\begin{align}
\tfrac{1}{12}&(c_{16}+c_{26}+c_{36})=(W^*_{34}+2W^*_{35})M_1M_2\notag \\[0.5ex]
&+(W^*_{36}+2W^*_{37})M_1'M_2'+\tfrac{1}{2}W^*_{38}(M_1M_2'+M_1'M_2).
\end{align}

For a decoupled model of the form
\begin{equation}
W^*=f(J)+\W(\I_1,\I_2,\I_4,\I_5,\I_6,\I_7,\I_8),\label{assumption8}
\end{equation}
it follows that $W^*_{34}=W^*_{35}=W^*_{36}=W^*_{37}=W^*_{38}=0$ and hence the Voigt constants must be interrelated according to
\begin{align} \label{special3}
& c_{11}+c_{12}-c_{33}-c_{23}=0, \notag \\ & c_{22}+c_{12}-c_{33}-c_{13}=0, \notag \\ & c_{16}+c_{26}+c_{36}=0,
\end{align}
and the 13 constants are reduced to 10.  Thus the material is not fully monoclinic in the linearized limit. It follows that materials with two families of non-orthogonal fibres for which the strain-energy function can be decomposed additively as \eqref{assumption8} do not behave like monoclinic solids when subject to infinitesimal deformations. It can also be checked that a monoclinic material for which the restrictions \eqref{special3} hold deforms in pure dilatation under hydrostatic stress.

By switching the indices 1 and 3 in the first two results in \eqref{special3} the results \eqref{special1} in Section \ref{Inconsistency with orthotropy} are recovered (here $\vec{M}$ and $\vec{M}'$ define the $(x_1,x_2)$ plane whereas in Section \ref{Inconsistency with orthotropy} they define the $(x_2,x_3)$ plane).


\section*{acknowledgements}
This work was supported by the Royal Society through an International Joint Project awarded to the second and third authors.
Finally the authors are grateful to Jerry Murphy (Dublin City University) for stimulating discussions on the topic.



\end{document}